\documentclass[aps,prb,twocolumn,superscriptaddress,floatfix]{revtex4}
\usepackage{psfig}

\bibstyle{apsrev.bib}

\begin{document}
\input epsf.sty

\title{Out-of-equilibrium critical dynamics at surfaces: Cluster dissolution and non-algebraic correlations}

\author{Michel Pleimling}
\affiliation{Institut f\"ur Theoretische Physik I, Universit\"at
Erlangen-N\"urnberg, D-91058 Erlangen, Germany}

\author{Ferenc Igl\'oi}
\affiliation{
Research Institute for Solid State Physics and Optics,
H-1525 Budapest, P.O.Box 49, Hungary}
\affiliation{
Institute of Theoretical Physics,
Szeged University, H-6720 Szeged, Hungary}

\date{\today}

\begin{abstract}
We study nonequilibrium dynamical properties at a free surface after the system is quenched from the high-temperature phase into the critical point. We show that if the spatial surface correlations decay sufficiently rapidly the surface magnetization and/or the surface manifold autocorrelations has a qualitatively different universal short time behavior than the same quantities in the bulk. At a free surface cluster dissolution may take place instead of domain growth yielding stationary dynamical correlations that decay in a
stretched exponential form. This phenomenon takes place in the three-dimensional Ising model and should be observable in real ferromagnets.
\end{abstract}

\maketitle

\newcommand{\bc}{\begin{center}}
\newcommand{\ec}{\end{center}}
\newcommand{\be}{\begin{equation}}
\newcommand{\ee}{\end{equation}}
\newcommand{\beqn}{\begin{eqnarray}}
\newcommand{\eeqn}{\end{eqnarray}}

Dynamical properties of systems which are quenched from the high-temperature phase to or below the critical point are of recent interest, both experimentally and theoretically (see Ref. \cite{bray,godr,ritort}
for reviews). In the bulk of the system, which is generally studied in this context, an ordering phenomenon takes place, which is manifested by the existence of a characteristic length-scale, $\xi(t)$, growing in time as $\xi(t) \sim t^{1/z}$, where $z$ is the well known dynamical exponent. If the quench is performed to the critical temperature new (nonequilibrium) dynamical exponents have to be defined. The short time dependence of the magnetization starting from a small finite initial value, $m_i$, is given by $m(t) \simeq m_i t^{\theta_b}$, where the initial slip exponent is expressed as\cite{jss89} $\theta_b=(x_i-x_b)/z$
. Here $x_b$ is the static scaling dimension of the bulk magnetization, whereas $x_i$ is a new nonequilibrium exponent, the scaling dimension of the initial magnetization. Since generally due to fluctuations $x_i > x_b$ there is a universal small-time increase of the order. The initial disordered state has a long-lasting effect on the autocorrelation function, too, measured at time $t$, after a waiting time $s<t$. In the limit $t \gg s$ the autocorrelation function behaves as\cite{huse89} $C(t,s) \sim t^{-\lambda_b/z}$, where the autocorrelation exponent in a $d$-dimensional system is given by $\lambda_b=d-x_i+x_b$.

Real materials are bounded by surfaces and many nonequilibrium processes (thermalization, transfer of heat, etc.)
are very intensive at the surface. At a free boundary layer the order is weaker than in the bulk due to missing bonds and surface correlations involve a new scaling dimension, $x_1>x_b$. Due to the inhomogeneity at
the surface the spatial and dynamical correlations are anisotropic: critical correlations between a surface point and one in the bulk has an exponent $x_1+x_b$ and only correlations between two surface points involve the exponent $2x_1$ \cite{binder,dd83}.

The phenomena of nonequilibrium dynamics have also been studied at surfaces by numerical
and by scaling and field-theoretical (FT) methods\cite{RiCz_95,ms96}. It is found that the scaling dimension of the initial magnetization, $x_i$, is the same as in the bulk, but the local magnetization operator has its surface dimension, $x_1$, instead of $x_b$. Therefore the exponent of the relaxation process, which involves one magnetization operator, is expected to change as $\theta_1=\theta_b-(x_1-x_b)/z$, whereas the autocorrelation of two operators will have an exponent\cite{RiCz_95} of $\lambda_1=\lambda_b+2(x_1-x_b)$. The FT results have been checked numerically, but the simulations were restricted to the two-dimensional (2d) and 3d Ising models.

In this Letter we are going to revisit the problem of nonequilibrium critical dynamics at surfaces. We show that during domain growth (DG) 
the correlated sites in the domain are either percolating (to which we refer as conventional DG dynamics) or they form isolated finite
clusters which gradually shrink as time goes on (in which case we use the term cluster dissolution (CD)), depending on the critical exponents of the given system. In
the CD regime relaxation and short time autocorrelations have unconventional, stretched exponential time
dependence, which are shown by scaling theory and checked by numerical calculations. This process takes
place in the 3d Ising model and should be experimentally observable in
real magnets. Surface manifold autocorrelations and persistence have also been studied.

In a phenomenological theory the number of correlated points within a domain of volume $V \sim \xi^d$ can be
estimated as the result of two competitive processes. i) Initial nonequilibrium growth creates correlations
among $V_i \sim \xi^{x_i}$ sites, ii) which are reduced by equilibrium critical fluctuations by a factor of $\xi^{-x_l}$. (Here we use the notation $x_l$ for the local exponent, which is $x_b$ in the bulk and $x_1$ at the surface.) If $x_i>x_l$, the correlated points in the domain form a percolating cluster, the mass of which scales as: $M \sim \xi^{x_i-x_l}$. In this DG regime the magnetization
at time $t$ is given by $m(t) \sim m_i M$ with the initial magnetization $m_i$. This formula recovers the known FT results, both in the bulk and at the surface. For $x_i<x_l$, however, the correlated points in the domain form only finite clusters and in this regime the CD dynamics applies.

The autocorrelation function in the long time limit, $s \gg 1$, $t-s \gg s$, is invariant under the
scale transformation, $C(t,s)=b^{-\eta} C(tb^{-z},sb^{-z})$, when lengths are rescaled as $L \to L/b$, $b>1$.
Here $\eta=2 x_l$ is the decay exponent of the local magnetization. With the choice $b=\xi_s=s^{1/z}$ we
arrive at the result:
\be
C(t,s)= c(s) f(t/s),\quad c(s) \sim \xi_s^{-\eta}, \quad t-s \gg s\;,
\label{C0}
\ee
where the pre-factor, $c(s)$, can be interpreted as the product of the critical magnetizations in two
cells of size $\xi_s$. In this interpretation the domain at time $t \gg s$, with typical size $\xi$, is divided into a large number of cells.
In the short time limit, $t-s<s$, however, when the domain contains only one cell, the spatial correlations
between two points of reference are in the DG regime of local-bulk type,
leading to the scaling form of the autocorrelation:
\be
C(t,s)= c_1(s) f_1[(t-s)/s],\quad c_1(s) \sim \xi_s^{-\eta_1}, \quad t-s < s\;,
\label{C1}
\ee
with $\eta_1=x_b+x_l$. Thus in an inhomogeneous environment the autocorrelation function
can be formally written as:
\be
C(t,s)=c(s)f(t/s)=\frac{c(t)}{c_1(t)} C_1(t,s),\quad t-s \gg s\;,
\label{C2}
\ee
where $C_1(t,s)$ is the autocorrelation in a homogeneous environment with local exponent $x_l$.
In the DG regime $C_1(t,s)$ follows from the observation that out of the volume of the domain there are $M$ sites which are correlated with the initial one, leading to
\be
C_1(t,s) \sim M/V \sim \xi^{-(d-x_i+x_l)}\;.
\label{C_b}
\ee
For a bulk site, with $x_l=x_b$ and $c(t)=c_1(t)$ we immediately obtain $C(t,s) \sim t^{-(d-x_i+x_b)/z}$ in
accordance with FT calculations. Similarly, for a surface point with $x_l=x_1$ and $c(t)/c_1(t) \sim
t^{(x_b-x_l)/z}$ we recover the FT result.

Let us now turn to the CD regime, when i) the cells contain only finite clusters and ii) in the short time
limit the two points of reference are within the same cell. In a given cell the probability to find a large cluster of linear size, $L$, is exponentially small: $P(L) \sim \exp(-a L^d)$, but the relaxation time, $t_r$, related to this cluster is large and can be estimated as follows. Correlations within the cluster are destroyed if a domain wall of $W \sim L^{d-1}$ sites are created within it. The number of still
correlated sites in the wall at time $t$ is estimated as $\sim W t^{(x_i-x_l)/z}$, from which the value of
the relaxation time, $t_r \sim L^{z(d-1)/(x_l-x_i)}$, follows. Now the dominant behavior of the autocorrelation function is obtained as an average over the clusters and is given by:
\beqn
C(\tau) & \sim & \int P(t_r) \exp(-\tau/t_r) {\rm d} t_r  \sim  \exp(-C \tau^\kappa),
\nonumber\\
\kappa & = & \frac{(x_l-x_i)d}{z(d-1)}\;.
\label{auto-}
\eeqn
in terms of $\tau=t-s < s$. If it happens that $\kappa>1$, then the decay is pure exponential. Note that the autocorrelation function is stationary
as long as the two points of reference are within the same cell. In the long time limit, $t \gg s$, however,
when the two points are in different cells, one can use a coarse-grained picture in which the DG theory,
with the results given below Eq.(\ref{C_b}), becomes applicable.

We can also estimate the short time relaxational behavior of the initial magnetization, $m_i$, by noting that
the dominant process in this case is the same as for the autocorrelation function. From this we obtain a fast stretched exponential short time decay of:
\be
m(t) \simeq  m_i \exp(- C_m t^\kappa), \quad t<t_i\;.
\label{m-}
\ee
with the universal exponent, $\kappa$, as given in Eq.(\ref{auto-}) and the decay is pure exponential if $\kappa>1$. At $t_i\sim |\ln m_i|^{-1/\kappa}$ the relaxation crosses over to a power-law behavior.

The above scaling results  are checked by numerical simulations of a 2d Ising model in which the strength of the coupling at a distance $y$ from the free surface deviates asymptotically from its bulk value by an amount of $A/y$. In this Hilhorst-van Leeuwen (HvL) model\cite{hvl}, the surface magnetization scaling dimension is known exactly\cite{ipt}: $x_1=(1-A/A_c)/2$, for $A<A_c$ and $A_c$ is a constant. The presence of the surface defect does not modify the value of the bulk exponents, which are known to be: $x_b=1/8$, $x_i=.53$ and $z=2.17$.

In the actual
simulation we considered $L \times L$ finite systems up to $L=300$ with periodic boundary condition (b.c.) in the
direction of translation invariance and free b.c. in the perpendicular, non-homogeneous direction and made
measurements up to several hundred time steps. The calculated
short time behavior of the surface magnetization is shown in Fig. 1a. It has indeed different characteristics for $x_i > x_1$ and for $x_i < x_1$, respectively. In the former case, the initial slip
behavior sets in after a short time period, whereas in the latter (CD) region there is a fast non-algebraic short time decrease of the magnetization (grey lines) before the much slower power-law decay starts. In the fast short time region the decay
correctly fits the predicted universal stretched exponential behavior of Eq. (\ref{m-}), as shown in the inset of Fig. 1a.

{
\begin{figure}[h]
\centerline{\epsfxsize=3.25in\ \epsfbox{
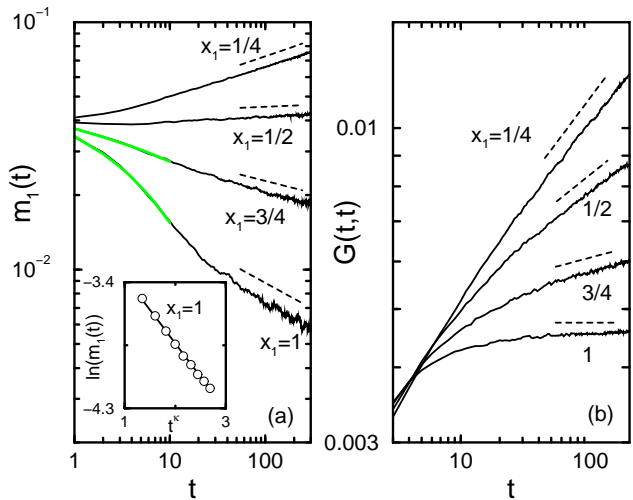}
}
\caption{
Short time behavior of (a) the surface magnetization and (b) the surface manifold autocorrelation
function $G(t,s)$ at the waiting time $t=s$ in the HvL model for various values of
the surface magnetization scaling dimension $x_1$. The broken straight lines indicate the predicted long time asymptotics. Here and in the following, the data, which
have been obtained by averaging over typically 30000 different runs, are free of
finite size effects. The inset in Fig.\ 1a shows the expected stretched exponential behavior
for $x_1=1$
}
\label{Abb1}
\end{figure}
}

In our next calculation we compared the short time behavior of the surface autocorrelation function in the 3d and
in the 2d Ising models. For the 3d Ising model with $x_b=.516$, $x_1=1.26$ (ordinary transition), $x_i=.74$ and $z=2.04$ we are in the CD regime, since $x_i < x_1$, whereas the (pure) semi-infinite 2d Ising model is in the DG regime. For the 3d model the simulations were performed on a cube of linear size up to $L=60$ with free b.c. in one direction and periodic b.c.-s in the other two directions.
The numerical results for 3d (Fig. 2a) show a stationary regime for $t-s<s$ in which the decay fits very well with the theoretical prediction in Eq.(\ref{auto-}) (see inset). In contrary for 2d (Fig. 2a) there is no
stationary regime, even not for very short time differences (see inset).

{
\begin{figure}[h]
\centerline{\epsfxsize=3.25in\ \epsfbox{
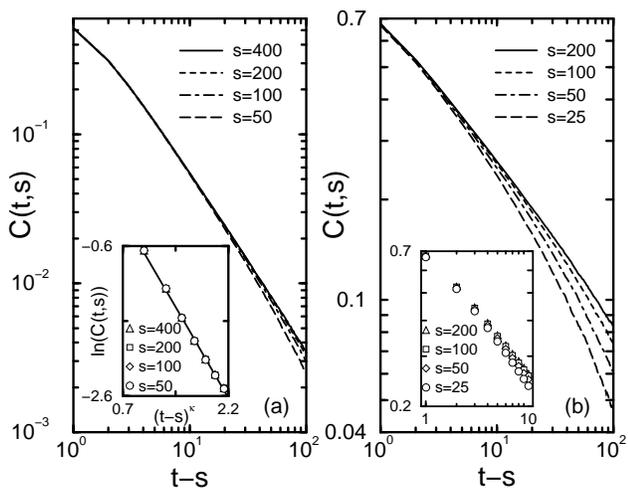}
}
\caption{
Surface autocorrelation function of (a) the 3d and (b) the 2d semi-infinite Ising models
as function of $t-s$. As shown in the insets, a stationary stretched exponential behavior
is observed in 3d with the exponent $\kappa$ in Eq.(\ref{auto-}), whereas in 2d no stationary short time regime is found.
}
\label{Abb2}
\end{figure}
}

We have also considered nonequilibrium dynamics related to a $d'$ dimensional
manifold, in particular in the case where the manifold is the surface layer of a semi-infinite system, thus $d'=d-1$.
The manifold autocorrelation function, $G(t,s)$, at the waiting time $t=s$ has a non-trivial behavior:
$G(t,t) \sim \xi^{(d'-x_l-x_b)}$, provided $d'_f=d'-x_l-x_b>0$. Thus in this case
the initially correlated points in the manifold form a percolating cluster inside the cell, with
a fractal dimension of $d'_f$. This situation corresponds to the DG process in this problem. If,
however $d'-x_l-x_b<0$, the manifold autocorrelations $G(t,t)$ approach a finite, time independent
value, thus the initially correlated sites within the growing cell are finite, isolated clusters,
which are then the subject of dissolution during the CD process. These predictions are successfully verified by numerical calculations on the 2d HvL model for different values of $x_1$ as shown in Fig. 1b.

For $t>s$ the considerations about $G(t,s)$ are simple in the DG region, i.e. when $d'-x_l-x_b>0$  and the result can be obtained by
integrating the related formulae for a single spin. For example we get from Eq.(\ref{C0}) for the asymptotic
behavior of the manifold autocorrelation function in the long time limit: $G(t,s)=g(s) f'(t/s)$, where $g(s) \sim  s^{(d'-2x_l)/z}$. Thus the waiting time dependence is a subject of change as time goes on, and the scaling combination: $s^{(x_l-x_b)/z}G(t,s)/G(s,s)$ is just the function of $t/s$, as illustrated in Fig. 3
for the 2d HvL model. The time dependence of the manifold autocorrelation function in the long time limit is obtained from Eq.(\ref{C2}) as $G(t,s) \sim t^{-\lambda'/z}$. Here $\lambda'=d-d'-x_i-x_b+2x_l$, from which both the bulk ($x_l=x_b$) and the surface
($x_l=x_1$) results follow.

{
\begin{figure}[h]
\centerline{\epsfxsize=3.25in\ \epsfbox{
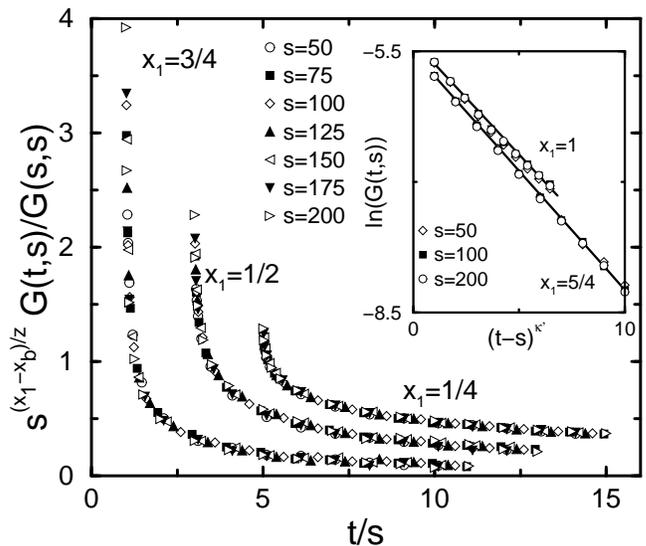}
}
\caption{
Rescaled surface manifold autocorrelation as function of $t/s$ in the DG regime of the
HvL model. For illustrative reasons the data points for $x_1=1/2$ resp.\ $x_1=1/4$
have been shifted horizontally by 2 resp.\ 4 units. The expected stretched (pure) exponential
behavior in the CD regime is shown in the inset.
}
\label{Abb3}
\end{figure}
}

In the CD region, when $d'-x_l-x_b<0$, $G(s,s)$ has a finite value, thus in the cell of size $\xi_s$ there are only finite clusters. In the short time limit, $t-s<s$, as for the single spin case the two points of reference
are within the same cell and the autocorrelations are governed by the CD dynamics.
In this case the driving force of the
dissolution is the inhomogeneity effect as local-bulk correlations go in time over to local-local correlations, as seen in Eqs.(\ref{C0}) and (\ref{C1}) for a single spin. In a time period, $\tau=t-s<s$, a cluster shrinks by a factor of
$\sim \tau^{(x_b-x_l)/z}$. Now using the same reasoning as for the single spin problem we arrive to the same conclusion, just $x_i$ in the single site result
has to be replaced by $x_b$ for a manifold. Thus the surface manifold autocorrelation function in the short time
limit, $t-s<s$, is stationary and described by a stretched exponential form:
$G(t,s) \sim \exp(-C' (t-s)^{\kappa'})$, with $\kappa'=(x_1-x_b)d/(z(d-1))$, unless $\kappa'>1$, when the decay
is a pure exponential. Again, in the long time limit, $t-s \gg s$, when the two points of reference are in different cells in a coarse-grained picture we recover the conventional DG results.
Numerical calculations for the 2d HvL model are in complete agreement with the theoretical predictions.
As seen in the inset of Fig. 3 the decay is stationary and for $x_1=1$ it is stretched exponential, whereas
for $x_1=5/4$ it is pure exponential.

{
\begin{figure}[h]
\centerline{\epsfxsize=3.25in\ \epsfbox{
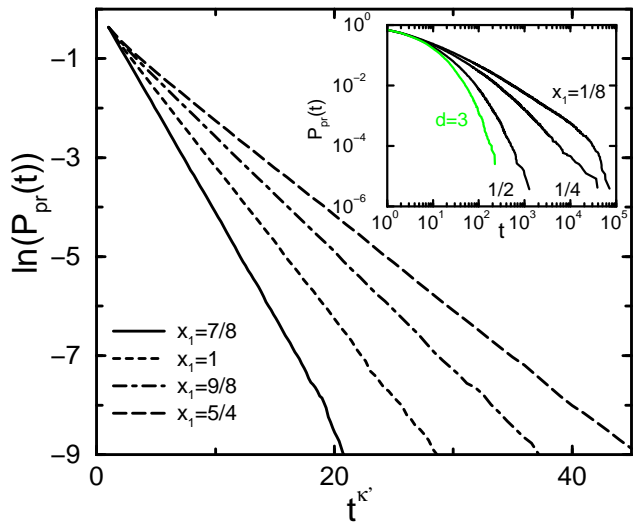}
}
\caption{
Time dependence of the surface manifold persistence. The main figure shows the stretched (pure)
exponential behavior in the regime $d'-x_l-x_b < 0$, whereas the power-law behavior
for $d'-x_l-x_b >0$ is displayed in the inset.
}
\label{Abb4}
\end{figure}
}

We close our Letter by presenting results concerning the surface manifold persistence, which completes recent studies
by Majumdar and Bray \cite{mb03} about persistence of manifolds which are embedded into the bulk. Since
persistence and (the short time behavior of) manifold autocorrelations are intimately related,  the persistence also shows qualitatively different behavior for $d'-x_l-x_b>0$ and $d'-x_l-x_b<0$, respectively. In the first (DG) regime
the persistence is in a power-law form: $P_{pr}(t) \sim t^{-\Theta_{pr}}$, where $\Theta_{pr}$ is some non-trivial
function of $d'_f=d'-x_l-x_b$. This result follows from a general theorem by Newell and Rosenblatt\cite{NR}.
On the other hand in the CD regime with $d'-x_l-x_b<0$ the persistence has a
faster, stretched exponential or pure exponential behavior, as the autocorrelation function: thus
$P_{pr}(t) \sim \exp(-a_1 t^{\kappa'})$, if $0<\kappa'<1$, otherwise $P_{pr}(t) \sim \exp(-a_2 t)$,
if $\kappa'>1$. We note that for a bulk manifold the persistence shows also these three different types of
functional forms, however the border between the different regions as well as the actual value of the exponent
in the stretched exponential are different in the two problems \cite{mb03}.

Numerical results in Fig. 4 for the 2d HvL model are in complete agreement with the predicted stretched
(pure) exponential decay in the regime $d'-x_l-x_b<0$. Results in the $d'-x_l-x_b>0$
regime are presented in
the inset to Fig. 4. The long-time power-law dependence is clearly seen for the different models, which include:
i) the surface manifold of the 3d Ising model, ii) the surface manifold of the 2d Ising model ($x_1=1/2$) and
iii) the case $x_1=x_b=1/8$ where we expect the persistence of a bulk manifold in the 2d Ising model.

In summary, we have presented a scaling theory describing nonequilibrium critical
dynamics at free surfaces. Depending on the values of the involved scaling dimensions, domain growth
or cluster dissolution is observed. The latter phenomenon, which has not been
observed up to now,
leads to unconventional behaviors of dynamical correlation functions. The theoretical predictions
have been verified in extensive Monte Carlo simulations.

We thank the Regionales Rechenzentrum Erlangen for the extensive use of the IA32 compute cluster.
This work has been supported by the Hungarian National
Research Fund under  grant No OTKA TO34183, TO37323,
MO28418 and M36803, by the Ministry of Education under grant No FKFP 87/2001,
and by the EC Centre of Excellence (No. ICA1-CT-2000-70029).

\end{document}